\documentclass[letter]{IEEEtran}
\IEEEoverridecommandlockouts
\usepackage{cite}
\usepackage{amssymb,amsfonts}
\usepackage[tbtags]{amsmath}
\usepackage{graphicx}
\usepackage{subcaption}
\usepackage{textcomp}
\usepackage{xcolor}
\usepackage{soul}
\usepackage{url}
\usepackage{stfloats}
\usepackage{makecell}
\usepackage{mathtools}
\usepackage{float}
\usepackage{gensymb}
\usepackage{booktabs}
\usepackage{bm}
\usepackage{algorithm}
\usepackage{algpseudocode}
\usepackage[export]{adjustbox}
\def\BibTeX{{\rm B\kern-.05em{\sc i\kern-.025em b}\kern-.08em
    T\kern-.1667em\lower.7ex\hbox{E}\kern-.125emX}}

\usepackage{hyperref}
\hypersetup{
    colorlinks=true,
    linkcolor=blue,
}

\begin{document}


\bstctlcite{IEEEexample:BSTcontrol}

\title{On the Spectral Efficiency of Movable and Rotary Antenna Arrays under Rician Fading}

\author{Eduardo N. Tominaga, \textit{Student Member, IEEE}, Onel L. A. López, \textit{Senior Member, IEEE}, Tommy Svensson, \textit{Senior Member, IEEE}, Richard D. Souza, \textit{Senior Member, IEEE}, Hirley Alves, \textit{Member, IEEE}
\thanks{
This research was financially supported by Research Council of Finland (former Academy of Finland), 6Genesis Flagship (grant no. 346208), European union’s Horizon 2020 research and innovation programme (EU-H2020), Hexa-X-II (grant no. 101095759) project, the Finnish Foundation for Technology Promotion, and in Brazil by CNPq (305021/2021-4, 402378/2021-0) and RNP/MCTIC 6G Mobile Communications Systems (01245.010604/2020-14). (\textit{Corresponding author: Eduardo N. Tominaga})

Eduardo N. Tominaga, Onel L. A. López, and Hirley Alves are with the Centre for Wireless Communications (CWC), University of Oulu, Finland. (E-mail: \{eduardo.noborotominaga,onel.alcarazlopez,hirley.alves\}@oulu.fi).

Tommy Svensson is with the Department of Electrical Engineering, Chalmers University of Technology, 412 96 Gothenburg, Sweden (E-mail: tommy.svensson@chalmers.se).

Richard Demo Souza is with the Department of Electrical and Electronics Engineering, Federal University of Santa Catarina (UFSC), Florian\'{o}polis, 88040-370, Brazil. (E-mail: richard.demo@ufsc.br).
}
}

\maketitle

\begin{abstract}
Most works evaluating the performance of Multi-User Multiple-Input Multiple-Output (MU-MIMO) systems consider Access Points (APs) with fixed antennas, that is, without any movement capability. Recently, the idea of APs with antenna arrays that are able to move have gained traction among the research community. Many works evaluate the communications performance of Movable Antenna Arrays (MAAs) that can move on the horizontal plane. However, they require a very bulky, complex and expensive movement system. In this work, we propose a simpler and cheaper alternative: the utilization of Rotary Antenna Arrays (RAA)s, i.e. antenna arrays that can rotate. We also analyze the performance of a system in which the array is able to both move and rotate. The movements and/or rotations of the array are computed in order to maximize the mean per-user achievable spectral efficiency, based on estimates of the locations of the active devices and using particle swarm optimization. We adopt a spatially correlated Rician fading channel model, and evaluate the resulting optimized performance of the different setups in terms of mean per-user achievable spectral efficiencies. Our numerical results show that both the optimal rotations and movements of the arrays can provide substantial performance gains when the line-of-sight components of the channel vectors are strong. Moreover, the simpler RAAs can outperform the MAAs when their movement area is constrained.
\end{abstract}


\section{Introduction}

\par Multi-User Multiple-Input Multiple-Output (MU-MIMO) technologies play a crucial role in contemporary wireless communication networks such as 4G LTE Advanced \cite{lim2013}, 5G NR \cite{boccardi2014} and WiFi 6 \cite{avallone2021}. In MU-MIMO networks, a base station or Access Point (AP) equipped with multiple antennas serve multiple active devices at the same time. By utilizing beamforming techniques, MU-MIMO provides numerous benefits, including diversity and array gains, spatial multiplexing capabilities, and interference suppression. These benefits collectively enhance the capacity, reliability, and coverage of wireless networks \cite{heath2018}.

\par The vast majority of works investigating the performance of MU-MIMO networks consider APs equipped with Fixed Antenna Arrays (FAAs), i.e., antenna arrays with no movement capabilities. Nevertheless, the idea of antenna arrays that are able to move has gained attention in the research community\footnote{Preliminary results of this work were published in the conference version \cite{tominaga2024}. In that work, we studied only rotary antenna arrays. Since the optimization problem studied in that work has only one variable (we optimize the angular position of only a single AP), brute force search was used instead of PSO.} \cite{lozano2021,onel_francisco_2021,onel2021,zubow2023,lin2023,tominaga2024,zhu2023,xiao2023,zhu2023_2,zhu2024_1,zhu2024_2,ning2024}. The goal of moving and/or rotating antennas is to take advantage of spatially varying channel conditions within a confined space. By adjusting the position and/or orientation of the antenna, better channel conditions can be achieved. In Machine-Type Communication (MTC) scenarios, devices are typically deployed at fixed locations or exhibit low mobility, and the surrounding propagation environment changes slowly. Under these conditions, narrowband MTC offers limited time and frequency diversity, which restricts the potential for improving data rates and transmission reliability. In such cases, antennas with movement capabilities present a promising solution for achieving higher spatial diversity gains.

\subsection{Related Works}

\par The utilization of antenna arrays with movement capabilities is not something new. For instance, the authors in \cite{li2018} proposed a Direction of Arrival (DOA) estimation method that utilizes a rotary Uniform Linear Array (ULA) of antennas. The rotary ULA presents satisfactory performance for under-determined DOA estimations, where the number of source signals can be larger than the number of receive antenna elements. The performance of point-to-point Line-of-Sight (LoS) links where both the transmitter and receiver are equipped with a rotary ULA was studied in \cite{lozano2021}. Their setup is able to approach the LoS capacity at any desired Signal-to-Noise Ratio (SNR). López et. al. \cite{onel_francisco_2021,onel2021} and Lin et. al. \cite{lin2023} proposed the utilization of rotary ULAs for wireless energy transfer. They studied a system where a power beacon, or AP, equipped with a rotary ULA is constantly rotating and transmitting energy signals in the downlink to several devices. The devices harvest energy from the transmitted signal in order to recharge their batteries. The authors in \cite{zubow2023} developed and tested a prototype for hybrid mechanical-electrical beamforming for mmWave WiFi. Their experimental results in a point-to-point setup showed that the optimal rotation of the antenna array can bring significant improvements in throughput for both LoS and Non-LoS (NLoS) scenarios.

\par More recently, movable antennas, which are antennas that are able to move along the horizontal plane within a constrained movement area, have been proposed \cite{zhu2023,xiao2023,zhu2023_2,zhu2024_1,zhu2024_2}. By exploiting the wireless channel spatial variation in a confined region, the position of the antennas can be dynamically changed to obtain better channel conditions and improve the communication performance. However, their major drawback is their difficult implementation: each movable antenna requires two servo-motors, cables and slide tracks in order to move, which represents high deployment, operation and maintenance costs.

\par The authors in \cite{zhu2023_2} investigated how the extra degrees-of-freedom obtained with a movable antenna array can be combined with the beamforming vector to obtain full array gain over the desired direction and null-steering over all undesired directions. In \cite{zhu2024_1}, the authors consider a point-to-point wireless communication system in which both the transmitter and the receiver are equipped with an antenna that can move on a plane. Their numerical results show that the movement of the antennas can significantly increase the SNR of the wireless link in comparison with fixed antennas. In \cite{zhu2024_2}, the authors considered the uplink of a MU-MIMO system in which multiple user terminals are equipped with a single movable antenna (which can also move on a plane), while the BS is equipped with an FAA. They showed that the optimal antenna positions can minimize the total transmit power of the users subject to a minimum achievable rate requirement for each user.

\par UAVs operating as flying base stations \cite{wang2019,amponis2022} can be also interpreted as APs with movement capabilities. Their most notable advantage is the several degrees of freedom for movement, since a UAV can be positioned at any point of the coverage area, at any height, and their position can be easily changed as needed. However, the main drawbacks of their utilization are their limited load capacity, very high power consumption, and the consequent need for frequent recharges~\cite{mohsan2023}.

\par In our previous works \cite{tominaga2024,tominaga2024_OJCOMS}, we investigated rotary antenna arrays in a single AP setup \cite{tominaga2024} and in different distributed MIMO setups \cite{tominaga2024_OJCOMS}. Both works showed that, under Rician channels, where the Rician factor is high, the optimal rotation of the antenna arrays brings substantial gains on the Spectral Efficiency (SE). However, both works did not compare the rotary antennas with other alternatives such as the movable antennas studied in \cite{zhu2023,xiao2023,zhu2023_2,zhu2024_1,zhu2024_2}.

\par In addition to rotary, movable, and flying antennas, Ning et. al. \cite{ning2024} discussed other possible implementations such as sliding, foldable, and even liquid antenna arrays. They emphasized that the research field of antennas with movement capabilities is still in its infancy, and that there is an urgent need to explore new and cost-effective implementations that achieves a good balance between communications performance and implementation complexity.
Note that the most important advantages of the rotary ULAs when compared to the alternative approaches are the lower deployment, operation and maintenance costs, since each AP requires a single servo-motor to perform the rotation of its ULA~\cite{zubow2023,lin2023}.

\begin{table}[t!]
    \centering
    \caption{List of Acronyms}
    \begin{tabular}{l l l l}
        \toprule
        \textbf{Acronym} & \textbf{Definition} & \textbf{Acronym} & \textbf{Definition} \\
        \midrule
        AP & Access Point & NLoS & Non-LoS \\
        CSI & \makecell[l]{Channel State\\Information} & PSO & \makecell[l]{Particle Swarm\\Optimization} \\
        FAA & \makecell[l]{Fixed Antenna\\Array} & RA & Random Access \\
        LoS & Line-of-Sight & RAA & \makecell[l]{Rotary Antenna\\Array} \\
        MAA & \makecell[l]{Movable Antenna\\Array} & SE & Spectral Efficiency \\
        MIMO & \makecell[l]{Multiple-Input\\Multiple-Output} & SINR & \makecell[l]{Signal-to-Interference\\-plus-Noise Ratio} \\
        MRAA & \makecell[l]{Movable and Rotary\\Antenna Array} & SNR & \makecell[l]{Signal-to-Noise\\Ratio} \\
        MTD & \makecell[l]{Machine-Type\\Device} & ULA & \makecell[l]{Uniform Linear\\Array} \\
        MU & Multi-User & ZF & Zero Forcing \\
        \bottomrule
    \end{tabular}    
    \label{tableAcronyms}
\end{table}

\subsection{Contributions and Organization of the Paper}

\par In this work, we propose Rotary Antenna Arrays (RAAs) as a low cost and low complexity alternative to the Movable Antennas Arrays (MAAs) studied in the literature. We also propose the combination of both techniques into Movable and Rotary Antenna Arrays (MRAAs). We consider an uplink data transmission scenario under spatially correlated Rician fading, and we adopt the mean per-user achievable SE as the performance metric. The optimal position and/or rotation of the antenna arrays is computed based on estimates of the locations of the active devices and using Particle Swarm Optimization (PSO). Our numerical results based on Monte Carlo simulations show that all the approaches yield substantial performance gains compared to the case of a static AP when the LoS component of the channel vectors is strong. Besides, if the movement area of the APs is not constrained, the movable and RAA present the best performance, followed by the MAA and then the RAA. Conversely, in the case of constrained movement areas or large coverage areas, the RAAs outperform the MAAs.

\begin{figure*}[t]
    \centering
    \begin{minipage}[b]{0.48\linewidth}
        \centering
        \includegraphics[scale=0.4]{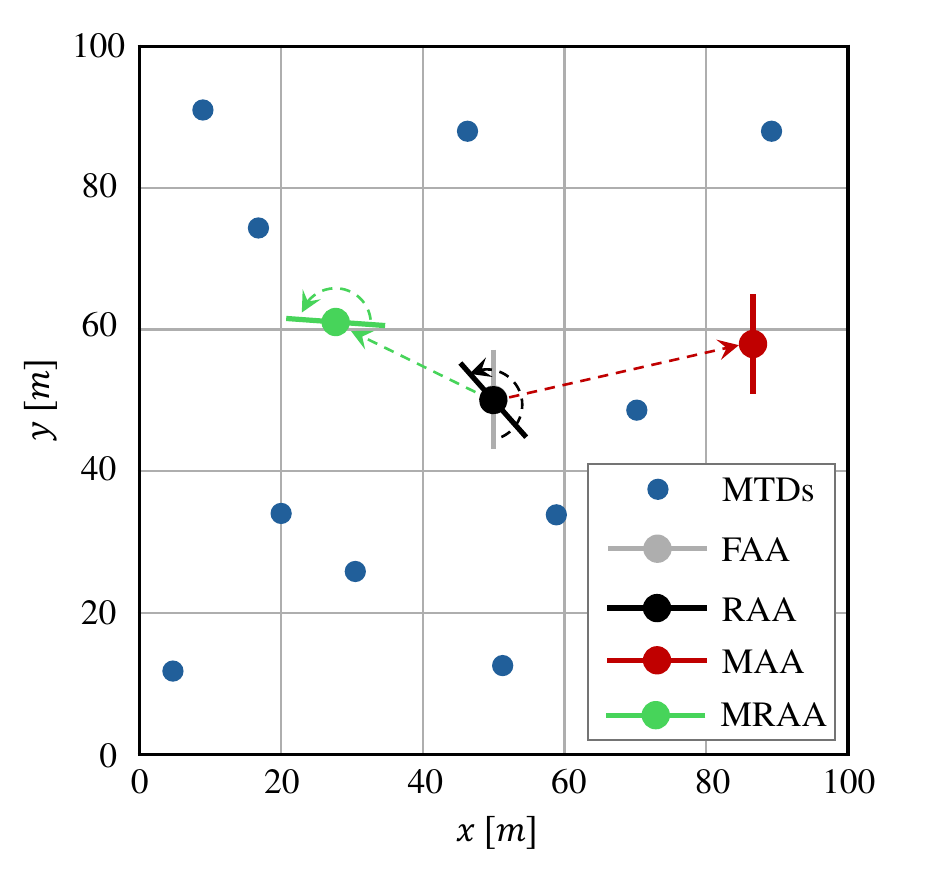}
        \caption{Illustration of the system model and the four types of APs considered in this work, for $L_A=100$~m and $K=10$.}
        \label{illustrationSystemModel}
    \end{minipage}
    \hspace{0.02\linewidth}
    \begin{minipage}[b]{0.48\linewidth}
        \centering
        \includegraphics[scale=0.45]{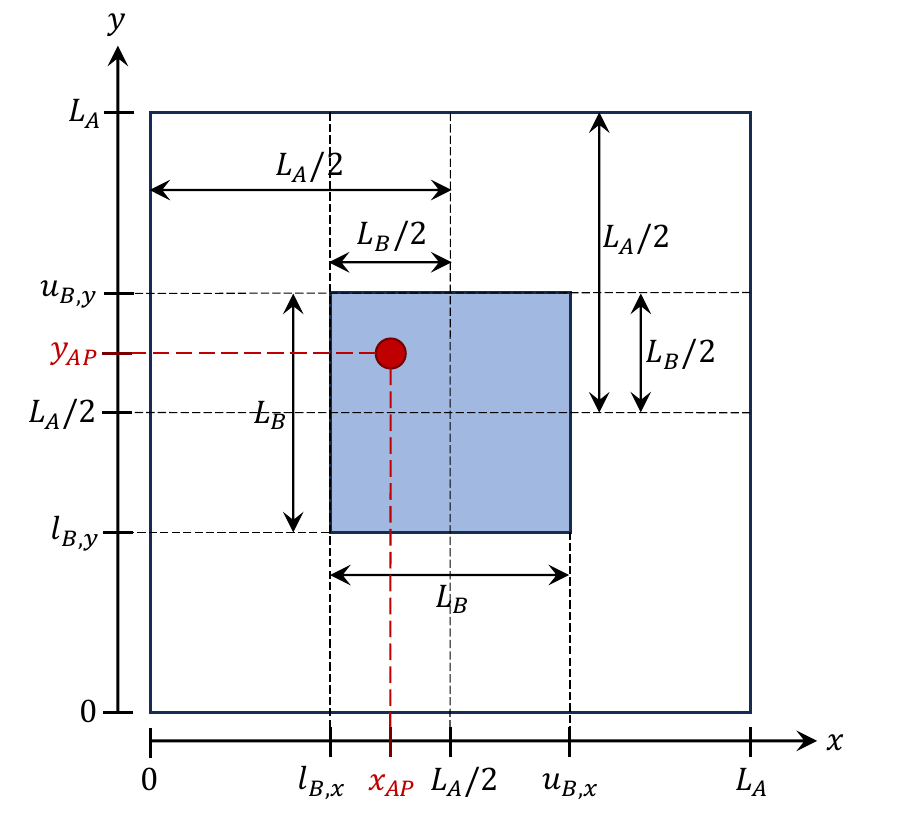}
        \caption{Illustration of the movement area of the MAAs (blue square). The red dot represents the position of the MAA.}
        \label{illustrationBoundaries}
    \end{minipage}
\end{figure*}

\par This paper is organized as follows. Section \ref{systemModel} presents the system and signal models, the proposed framework, the adopted performance metric and a mathematical model for the localization error. Section \ref{optimalAngularPositions} introduces the mechanism for the optimization of the angular position of the rotary ULAs and the proposed location-based beamforming method to compute the objective function. Section \ref{numericalResults} presents and discusses the numerical results. Finally, Section \ref{conclusions} concludes the paper. Table \ref{tableAcronyms} lists the acronyms used throughout this paper alphabetically.

\par \textbf{Notation:} lowercase bold face letters denote column vectors, while boldface upper case letters denote matrices. $a_i$ is the $i$-th element of the column vector $\textbf{a}$, while $\textbf{a}_i$ is the $i$-th column of the matrix $\textbf{A}$. $\textbf{I}_M$ is the identity matrix with size $M\times M$. The superscripts $(\cdot)^T$ and $(\cdot)^H$ denote the transpose and the conjugate transpose of a vector or matrix, respectively. The magnitude of a scalar quantity or the cardinality of a set are denoted by $|\cdot|$. The Euclidean norm of a vector (2-norm) is denoted by $\Vert\cdot\rVert$. We denote the one dimensional uniform distribution with bounds $a$ and $b$ by $\mathcal{U}(a,b)$. We denote the multivariate Gaussian distribution with mean $\mathbf{a}$ and covariance $\mathbf{B}$ by $\mathcal{N}(\mathbf{a},\mathbf{B})$.

\section{System Model}
\label{systemModel}

\par We consider an indoor\footnote{In an indoor scenario, the MAA can move on the ceiling of the coverage area. In an alternative outdoor scenario, the MAA would be able to move on the top or on the façade of a building.} square coverage area with dimensions $L_A\times L_A\;\text{m}^2$. The coverage area is served by a single Access Point (AP) equipped with a Uniform Linear Array (ULA) of $M$ half-wavelength spaced antenna elements, and placed at height $h_{\text{AP}}$. The AP is located at the center of the coverage area, i.e., $\textbf{p}_{\text{AP}}^0=(x_{\text{AP}}^0,y_{\text{AP}}^0)=(L_A/2,L_A/2)$.

\par The AP serves $K$ active Machine-Type Devices (MTDs) simultaneously. Let $\textbf{p}_k=(x_k,y_k)^T$ denote the coordinates of the $k$-th device, assuming for simplicity that all devices are located at the same height $h_{\text{device}}$ \cite{ngo2017,chen2018}.

\par
\begin{figure}[t!]
    \centering
    \begin{subfigure}{0.45\textwidth}

        \centering
        \includegraphics[scale=0.4]{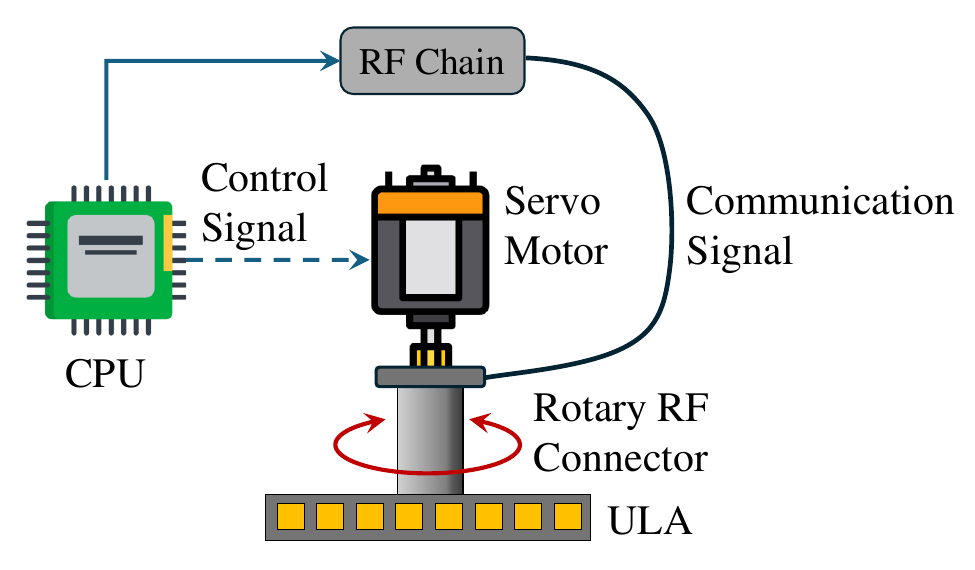}
        \caption{}
    \end{subfigure}
    \\
    \begin{subfigure}{0.45\textwidth}
        \centering
        \includegraphics[scale=0.32]{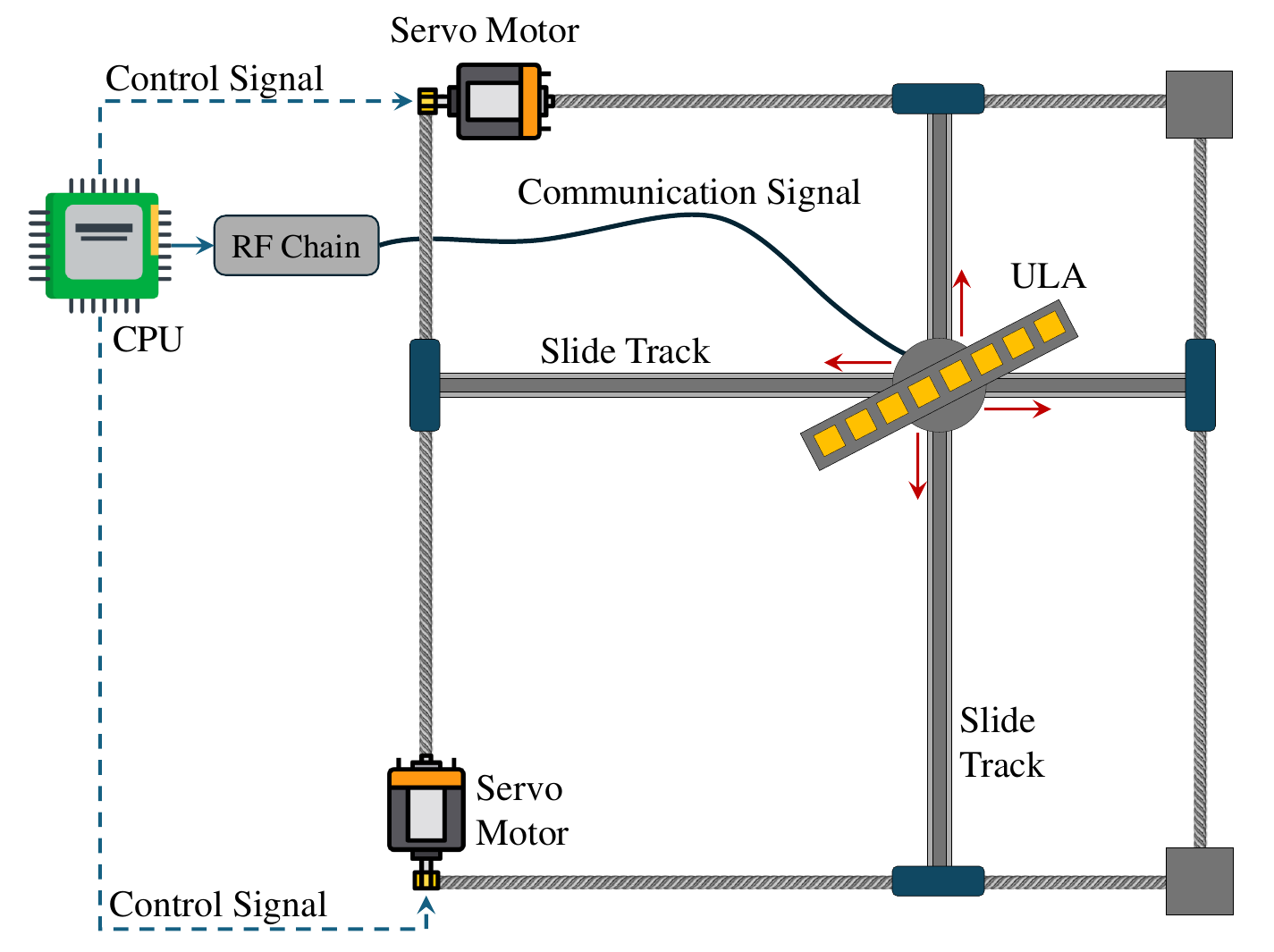}
        \caption{}
    \end{subfigure}
    \caption{Illustration of (a) RAA and (b) MRAA systems \cite{zhu2023}. The antenna array is an ULA of $M=8$ antenna elements.}
    \label{illustration_APs}
\end{figure}

 The AP can be equipped with one of four different types of antenna arrays:
\begin{enumerate}
    \item \textit{FAA}: the antenna array has no movement capabilities.
    \item \textit{RAA}: the antenna array is able to rotate.
    \item \textit{MAA}: the antenna array is able to move on the horizontal plane.
    \item \textit{MRAA}: the antenna array can both rotate and move on the horizontal plane.
\end{enumerate}
The system model is illustrated in Fig. \ref{illustrationSystemModel}. The MAAs and MRAAs are able to move within a square area with dimensions $L_B\times L_B$ that is inscribed on the square coverage area, as illustrated in Fig.~\ref{illustrationBoundaries}. Finally, Fig. \ref{illustration_APs} shows illustrations of the MAA and MRAA systems. Note that the RAA is equipped with a single servo motor, while the MAA has two servo motors, cables, and slide tracks. The MRAA has the same movement apparatus of the MAA, but equipped with an additional third servo motor that rotates the array.

\subsection{Channel Model}

\par In this subsection, we present a mathematical model that describes the wireless channels between the $K$ active MTDs and the ULA connected to the AP. Note that this model is applicable to any of the systems described previously.

\par We adopt a spatially correlated Rician fading channel model \cite{ozdogan2019}. Let $\textbf{h}_k\in\mathbb{C}^{M\times1}$ denote the channel vector between the $k$-th device and the AP. It can be modeled as \cite{dileep2021}
\begin{equation}
    \label{channelVector}
    \textbf{h}_k=\sqrt{\dfrac{\kappa}{1+\kappa}}\textbf{h}_k^{\text{los}} + \sqrt{\dfrac{1}{1+\kappa}}\textbf{h}_k^{\text{nlos}},
\end{equation}
where $\kappa$ is the Rician factor, $\textbf{h}_k^{\text{los}}\in\mathbb{C}^{M\times1}$ is the deterministic LoS component, and $\textbf{h}_k^{\text{nlos}}\in\mathbb{C}^{M\times1}$ is the random NLoS component.

\par The deterministic LoS component is given by
\begin{equation}
    \label{losComponent}
    \textbf{h}_k^{\text{los}}=\sqrt{\beta_k}
    \begin{bmatrix}
        1\\
        \exp(-j2\pi\Delta\sin(\phi_k))\\
        \exp(-j4\pi\Delta\sin(\phi_k))\\
        \vdots\\
        \exp(-j2\pi(M-1)\Delta\sin(\phi_k))\\
    \end{bmatrix},
\end{equation}
where $\beta_k$ is the power attenuation due to the distance between the $k$-th device and the AP, $\Delta$ is the normalized inter-antenna spacing, and $\phi_k\in[0,2\pi]$ is the azimuth angle relative to the boresight of the ULA of the AP. Meanwhile, the random NLoS component is distributed as
\begin{equation}
    \textbf{h}_k^{\text{nlos}}\sim\mathcal{CN}(\textbf{0},\textbf{R}_k).
\end{equation}
Note that
\begin{equation}
    \textbf{h}_k\sim\mathcal{CN}\left(\sqrt{\dfrac{\kappa}{1+\kappa}}\textbf{h}_k^{\text{los}},\dfrac{\textbf{R}_k}{\kappa+1}\right),
\end{equation}
where $\textbf{R}_k\in\mathbb{C}^{M\times M}$ with $\text{Tr}(\textbf{R}_k)=\beta_k$ is the positive semi-definite covariance matrix describing the spatial correlation of the NLoS components.

\par The spatial covariance matrices can be (approximately) modeled using the Gaussian local scattering model \cite[Sec. 2.6]{bjornson2017}. Specifically, the $s$-th row, $m$-th column element of the correlation matrix is
\begin{equation}
\begin{split}
    [\textbf{R}_k]_{s,m}=\dfrac{\beta_k}{N}\sum_{n=1}^N\exp[j\pi(s-m)\sin(\psi_{k,n})] \\
    \times \exp\left\{-\dfrac{\sigma_\phi^2}{2}[\pi(s-m)\cos(\psi_{k,n})]^2 \right\},
\end{split}
\end{equation}
where $N$ is the number of scattering clusters, $\psi_{k,n}$ is the nominal Angle of Arrival (AoA) for the $n$-th cluster, and $\sigma_\psi$ is the Angular Standard Deviation (ASD).

\par To take advantage of its multiple antennas, the AP needs to estimate the channel responses from the MTDs that are active. The channel estimation is often done using pilot sequences that the UEs transmit in the uplink and are known by the AP \cite{bjornson2017}. In practice, the channel estimates are not perfect, i.e., there is a channel estimation error associated to them. The estimated channel vector of the $k$-th device, $\Hat{\textbf{h}}_k\in\mathbb{C}^{M\times1}$, can be modeled as the sum of the true channel vector plus a random error vector as \cite{wang2012,eraslan2013,onel2022}
\begin{equation}
    \label{estimatedChannelVectors}
    \hat{\textbf{h}}_k=\textbf{h}_k+\Tilde{\textbf{h}}_k,
\end{equation}
where $\Tilde{\textbf{h}}_k\sim\mathcal{CN}(\textbf{0},\sigma_{\text{csi}}^2\textbf{I})$ is the vector of channel estimation errors. Note that the true channel realizations and the channel estimation errors are uncorrelated.

\par The parameter $\sigma_{\text{csi}}^2$ indicates the quality of the channel estimates. Let
\begin{equation}
    \rho=\dfrac{p}{\sigma_n^2}
\end{equation}
denote the per-AP antenna transmit SNR, where $p\geq0$ is the fixed uplink transmit power (which is the same for all the devices) and $\sigma_n^2$ is the receive noise power at the APs. We assume there are $\tau_p$ orthogonal pilot sequences during the uplink data transmission~phase, such that $\tau_p\geq K$. We also assume that the duration of the uplink pilot transmission phase is equal to $\tau_p$ symbols. Then, variance of the channel estimation errors can be modeled as a decreasing function of $\rho$ as \cite{wang2012,eraslan2013,onel2022}
\begin{equation}
    \sigma_{\text{csi}}^2=\dfrac{1}{\tau_p\rho}.
\end{equation}
Note that the channel estimation error depends only on the uplink transmit power, receive noise power and number of orthogonal pilots, thus it is the same for all devices.

\subsection{Signal Model}

\par The matrix $\textbf{H}\in\mathbb{C}^{M\times K}$ containing the channel vectors of the $K$ devices transmitting their data to the AP can be written~as
\begin{equation}
    \textbf{H}=[\textbf{h}_1,\textbf{h}_2,\ldots,\textbf{h}_K].
\end{equation}
Then, the $M\times 1$ received signal vector can be written as
\begin{equation}
    \textbf{y}=\sqrt{p}\textbf{H}\textbf{x}+\textbf{n},
\end{equation}
where $\textbf{x}\in\mathbb{C}^{K\times 1}$ is the vector of symbols simultaneously transmitted by the $K$ devices, and $\textbf{n}\in\mathbb{C}^{M\times 1}$ is the vector of additive white Gaussian noise samples such that $\textbf{n}\sim\mathcal{CN}(\textbf{0}_{M\times1},\sigma^2_n\textbf{I}_M)$.

\par Let $\textbf{V}\in\mathbb{C}^{M\times K}$ be a linear detector matrix used for the joint decoding of the signals transmitted from the $K$ devices. The received signal after the linear detection operation is split to $K$ streams and given by
\begin{equation}
    \textbf{r}=\textbf{V}^H\textbf{y}=\sqrt{p}\textbf{V}^H\textbf{H}\textbf{x}+\textbf{V}^H\textbf{n}.
\end{equation}
Let $r_k$ and $x_k$ denote the $k$-th elements of $\textbf{r}$ and $\textbf{x}$, respectively. Then, the received signal corresponding to the $k$-th device can be written as
\begin{equation}
    \label{r_k}
    r_k=\underbrace{\sqrt{p}\textbf{v}_k^H\textbf{h}_kx_k}_\text{Desired signal} + \underbrace{\sqrt{p}\textbf{v}_k^H\sum_{k'\neq k}^K \textbf{h}_{k'}x_{k'}}_\text{Inter-user interference} + \underbrace{\textbf{v}_k^H\textbf{n}}_\text{Noise},
\end{equation}
where $\textbf{v}_k$ and $\textbf{h}_k$ are the $k$-th columns of the matrices $\textbf{V}$ and $\textbf{H}$, respectively. From (\ref{r_k}), the signal-to-interference-plus-noise ratio of the uplink transmission from the $k$-th device is given by
\begin{equation}
    \label{gamma_k}
    \gamma_k=\dfrac{p|\textbf{v}_k^H\textbf{h}_k|^2}{p\sum_{k'\neq k}^K |\textbf{v}_k^H\textbf{h}_{k'}|^2+\sigma^2_n\lVert\textbf{v}_k^H\rVert^2}.
\end{equation}

\par The receive combining matrix $\textbf{V}$ is computed as a function of the matrix of estimated channel vectors $\hat{\textbf{H}}\in\mathbb{C}^{M\times K}$, $\hat{\textbf{H}}=[\hat{\textbf{h}}_1,\ldots,\hat{\textbf{h}}_K]$. In this work, we adopt Zero Forcing (ZF) combining\footnote{MMSE combining is the optimal linear receive combining scheme. However, its implementation requires statistical knowledge of the noise and interference \cite{bjornson2017}. Besides, the performance difference between ZF and MMSE is negligible in the high SNR regime \cite{tse2005}.}. The receive combining matrix is computed as \cite{liu2020}
\begin{equation}
    \label{combiningMatrix}
    \textbf{V}=\hat{\textbf{H}}(\hat{\textbf{H}}^H\hat{\textbf{H}})^{-1}.
\end{equation}

\subsection{Proposed Framework}

\par In this subsection, we describe our proposed framework for the optimization of the rotation and/or position of the antenna array and uplink data transmission. Inspired by the three-phase scheduled random access scheme from \cite{kang2022}, our framework has the following phases:
\begin{enumerate}
    \item Active MTDs, i.e., MTDs seeking to send data to the network, transmit non-orthogonal uplink pilots for activity detection.
    \item The AP identifies the set of active MTDs and, utilizing some indoor localization techniques, obtains estimates of the locations of the active MTDs.
    \item The AP assumes pure LoS propagation and utilizes the estimated locations of the MTDs and location based beamforming to compute its optimal rotation and/or position.
    \item The AP broadcasts a common downlink feedback message to assign each user an orthogonal pilot sequence.
    \item The MTDs transmit their orthogonal pilot sequences and data during multiple time slots. The uplink orthogonal pilots are used to compute the CSI estimates shown in (\ref{estimatedChannelVectors}) for each coherence time interval, which are then used to compute the ZF receive combining vectors in (\ref{combiningMatrix}). 
\end{enumerate}

\par
\begin{figure*}[t]
    \centering
    \includegraphics[scale=0.5]{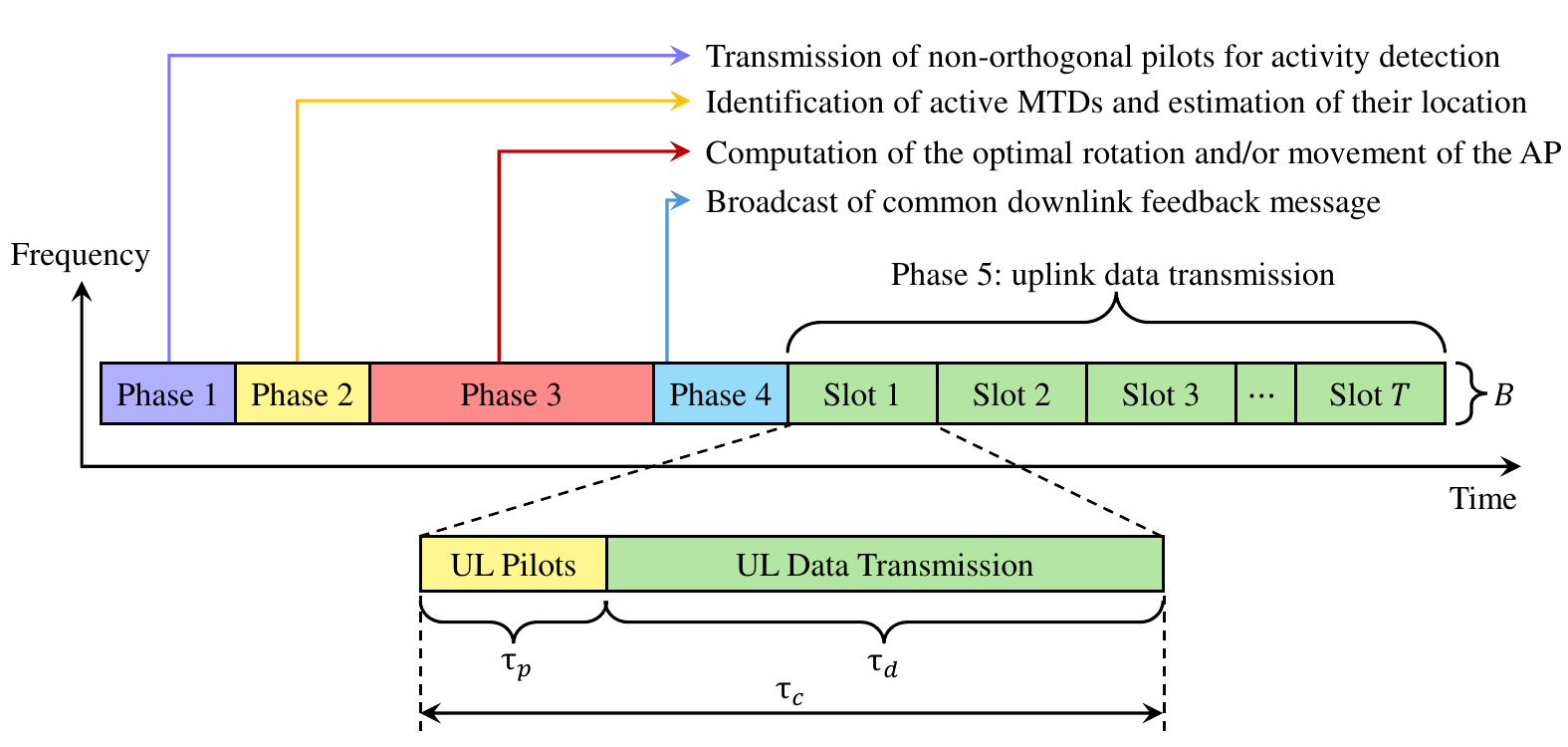}
    \caption{Illustration of the proposed framework for optimization of the rotation and/or position of the AP and uplink data transmission.}
    \label{illustrationProtocol}
\end{figure*}

 The proposed framework is illustrated in Fig. \ref{illustrationProtocol}. Note that phase 5 extends over $T$ time slots. Within each time slot, the $K$ active devices transmit simultaneously. Each time slot is as long as the channel's coherence time interval and contains $\tau_c$ symbols. The first $\tau_p$ symbols of each time slot are used for the transmission of the orthogonal pilot sequences, while the remaining $\tau_d=\tau_c-\tau_p$ symbols are utilized for uplink data transmission. The numerical results presented in this work, that is, the mean per-user achievable SE presented in Section \ref{performanceMetrics}, correspond to the uplink communication performance achieved in the data transmission.

\subsection{Performance Metrics}
\label{performanceMetrics}

\par We adopt as the performance metric the per-user mean achievable uplink Spectral Efficiency (SE). The achievable uplink SE of the $k$-th device is \cite{liu2020}
\begin{equation}
    \label{per-user-achievable-SE}
    R_k=\dfrac{\tau_d}{\tau_c}\mathbb{E}_{\textbf{H}}\{\log_2(1+\gamma_k)\}.
\end{equation}
The expectation is taken with respect to several realizations of the channel matrices $\textbf{H}$, which includes the effects of both LoS and NLoS propagation.

\par Then, the per-user mean achievable uplink SE is obtained by averaging over the achievable uplink SE of the $K$ devices,~i.e.,
\begin{equation}
    \label{mean-per-user-achievable-SE}
    \Bar{R}=\dfrac{1}{K}\sum_{k=1}^K R_k.
\end{equation}

\subsection{Localization Error Model}
\label{positioningErrorModel}

\par We adopt the same localization error model that was utilized in our previous work \cite{tominaga2024}. Considering that all the devices are at the same height $h_{\text{device}}$, the imperfect positioning impairment refers to the uncertainty on the location of the devices only on the $(x,y)$ directions. Let $\textbf{p}_k=(x_k,y_k)$ denote the true location of the $k$-th device, and $\hat{\textbf{p}}_k=(\hat{x}_k,\hat{y}_k)$ denote the estimated location. The localization error vector associated to the $k$-th device can be modeled as
\begin{equation}
   \textbf{e}_k=\textbf{p}_k-\hat{\textbf{p}}_k=(x_{e,k},y_{e,k}),
\end{equation}
where
\begin{align}
    & x_{e,k}=x_k-\hat{x}_k,\\
    & y_{e,k}=y_k-\hat{y}_k
\end{align}
are the $x$ and $y$ components of the localization error vector, respectively.

\par Aiming at generality (that is, in order to not introduce any additional assumption or bias related to indoor localization in our system model), we model the localization error as a bivariate Gaussian distribution, since it is the least informative distribution for any given variance \cite{gustafsson2005}. The localization error has mean $\bm{\mu}=[\textbf{0}\;\textbf{0}]^T$ and covariance matrix $\bm{\Sigma}=\sigma_e^2\textbf{I}_2$ \cite{zhu2022}. Then, the $x$ and $y$ components of the localization error vector follow a Normal distribution:
\begin{equation}
    x_{e,k},y_{e,k}\sim\mathcal{N}(0,\sigma_e^2).
\end{equation}

\par Note that the localization error does not affect the mean per-user achievable SE directly. In the case of FAAs, the imperfect information about the locations of the active MTDs has no effect. In the case of RAAs, MAAs and MRAAs, the imperfect location estimates affect the computation of the optimal rotation and/or position of the antenna array. By rotating/moving the array to a sub-optimal orientation/position, the potential gains on the SE obtained via the spatially varying channel conditions are diminished.

\begin{figure}[t!]
    \centering
    \begin{subfigure}{0.45\linewidth}
        \centering
        \includegraphics[scale=0.52]{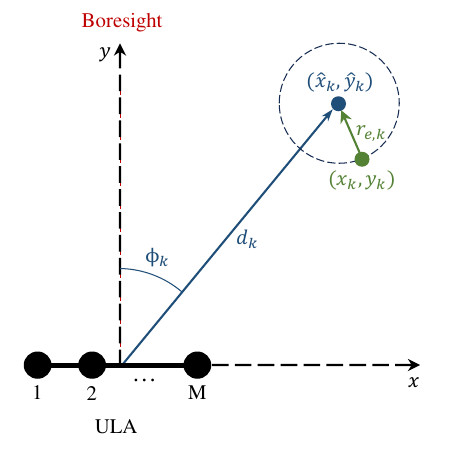}
        \caption{}
    \end{subfigure}
    \hspace{0.01cm}
    \begin{subfigure}{0.45\linewidth}
        \centering
        \includegraphics[scale=0.52]{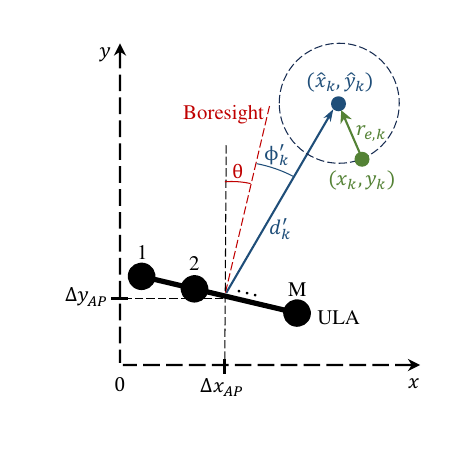}
        \caption{}
    \end{subfigure}
    \caption{Illustration of the ULA of an AP and its position and rotation with respect to one active MTD (a) before movements (b) after movements. The green circle represents the true location of the device, and the blue circle represents the estimated location.}
    \label{Illustration_Position_Angle}
\end{figure}

\section{Optimization of the Position and Rotation}
\label{optimalAngularPositions}

\par In each time slot, $K$ distinct devices are active. For each subset of $K$ locations of active devices, there is a distinct optimal rotation and/or optimal position for the antenna array.

\par Let $(x_{\text{AA}}^0,y_{\text{AA}}^0)$ denote the initial position of the array, and let $\theta_{\text{AA}}\in[0,\pi]$ denote its rotation. The array and its position with respect to an active MTD, before and after its movement and rotation, is illustrated in Fig. \ref{Illustration_Position_Angle}. The position of the array after its movement can be written as
\begin{align}
    & x_{\text{AA}}':=x_{\text{AA}}^0+\Delta x_{\text{AA}},\\
    & y_{\text{AA}}':=y_{\text{AA}}^0+\Delta y_{\text{AA}}.
\end{align}
The angle between the $k$-th device and the array after the rotation of the AP is given by
\begin{equation}
    \phi_k':=\phi_k+\theta_{\text{AA}}.
\end{equation}
These three variables can be jointly optimized. The optimization problem can be written as
\begin{equation}
    \begin{aligned}
    & \text{maximize} && f(x_{\text{AA}},y_{\text{AA}},\theta_{\text{AA}}\;|\;\Hat{\textbf{p}}_k,\forall k) \\
    & \text{subject to} && l_B \leq x_{\text{AA}}, y_{\text{AA}} \leq u_B, \\
    & && 0\leq\theta_{\text{AA}}\leq\pi,
    \end{aligned}
\end{equation}
where $f(\cdot)$ is the objective function to be maximized, and $l_B$ and $u_B$ are, the lower and upper bounds, respectively, for the movements of the array in both $x$ and $y$ directions. The mechanism utilized to obtain the objective function is described in Section \ref{locationBasedBeamforming}. Given the estimates of the locations of the active MTDs, $\textbf{p}_k\;\forall k\in\{1,\ldots,K\}$, we aim at obtaining the position and/or rotation of the array (that is, the variables $x_\text{AP}$, $y_\text{AP}$, and $\theta_{\text{AA}}$) that maximizes $f$. Considering that the MAAs can move only within a square with dimensions $L_B\times L_B$, inscribed on the square coverage area with dimensions $L_A\times L_A$, the lower and upper bounds for both $x_{\text{AA}},y_{\text{AA}}$ are respectively given by
\begin{align}
    & l_B=\dfrac{L_A-L_B}{2},\\
    & u_B=\dfrac{L_A+L_B}{2}.
\end{align}
Considering the symmetry of radiation pattern of the ULA, the rotation of the array is bounded in the range $\theta_{\text{AA}}\in[0,\pi]$~rad.

\subsection{Location-Based Beamforming}
\label{locationBasedBeamforming}

\par In this subsection, we present the mechanism utilized to compute the objective function $f(\cdot)$, which is maximized using PSO. Given the estimates of the locations of all active MTDs, that is, $\Hat{\textbf{p}}_k,\;\forall k$, we adopt a location-based beamforming \cite{maiberger2010,kela2016,yan2016,liu2016} approach to determine $f(\cdot)$ as a function of the position and rotation of the antenna array\footnote{We assume that the same set of active devices transmits data over multiple consecutive coherence time intervals. In this case, the optimization is performed solely based on the predicted LoS component of the channel vectors because it is deterministic, that is, it is constant over several coherence time intervals. If we perform the optimization of the rotation and/or position of the APs based on the estimated Rician channels, we would need to do it on every coherence time interval, which would be impractical considering the mechanical limitations of the proposed movable and rotary systems}.

\par Based on $\Hat{\textbf{p}}_k,\;\forall k$, the AP computes the estimates for the distances and azimuth angles between the AP and the MTDs, i.e. $\hat{d}_{k}$ and $\hat{\phi}_{k}$, $\forall k$. Then, it computes pseudo channel vectors assuming pure LoS propagation as
\begin{equation}
    \label{pseudoChannelVectors}
    \textbf{h}_k^{\text{pseudo}}=\sqrt{\Hat{\beta}_k}
    \begin{bmatrix}
        1\\
        \exp(-j2\pi\Delta\sin(\Hat{\phi}_k))\\
        \exp(-j4\pi\Delta\sin(\Hat{\phi}_k))\\
        \vdots\\
        \exp(-j2\pi(S-1)\Delta\sin(\Hat{\phi}_k))\\
    \end{bmatrix}.
\end{equation}
Note that the estimated large-scale fading coefficient $\Hat{\beta}_k$ is computed as a function of the estimated distances $\Hat{d}_k$ assuming a known channel model. Receive combining vectors are then computed as a function of the pseudo channel vectors according to (\ref{combiningMatrix}). Finally, the objective function is obtained by computing the predicted mean-per user achievable SE utilizing the pseudo-channel vectors from (\ref{pseudoChannelVectors}) and the corresponding receive combining vectors in (\ref{gamma_k}), (\ref{per-user-achievable-SE}), and (\ref{mean-per-user-achievable-SE}). The objective function corresponds to the predicted mean per-user achievable SE, which is predicted assuming assuming full LoS propagation, versus the rotation and/movement of the array\footnote{Note that this location-based beamforming approach, which relies on the pseudo-channel vectors from (\ref{pseudoChannelVectors}), is utilized only to compute the optimal rotation and/or position of the antenna array. The final performance metrics consider channel estimates that are obtained with pilot sequences transmitted in the uplink. The real mean per-user achievable SE is then computed using the estimated channel vectors in (\ref{gamma_k}) and (\ref{combiningMatrix}).}.

\begin{figure}[t]
    \centering
    \includegraphics[scale=0.55]{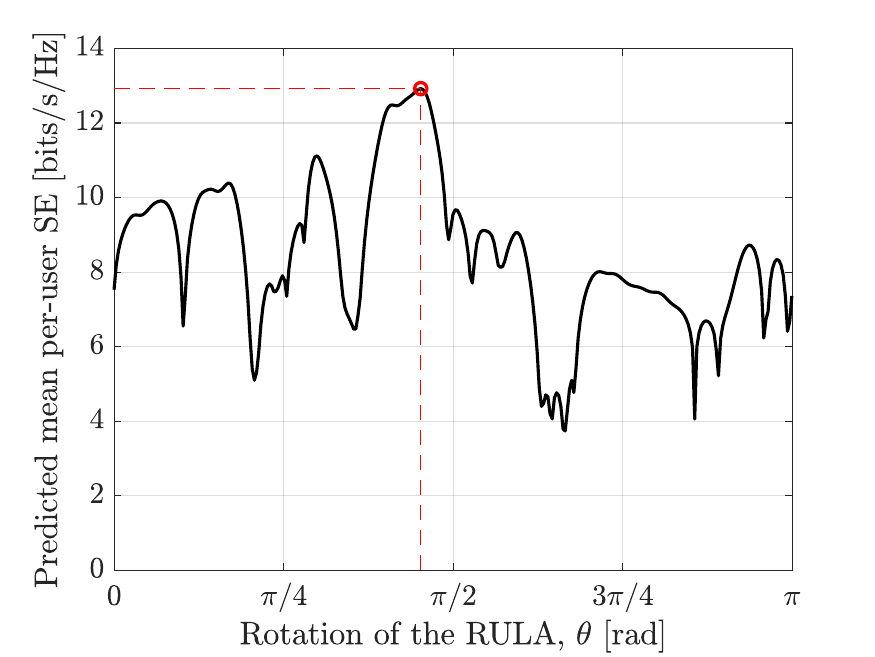}
    \caption{Objective function of the RAA, for $L_A=100$ m, $M=16$, and $K=10$.}
    \label{objectiveFunction_Rotary}
    \vspace{0.1cm}
    \includegraphics[scale=0.5]{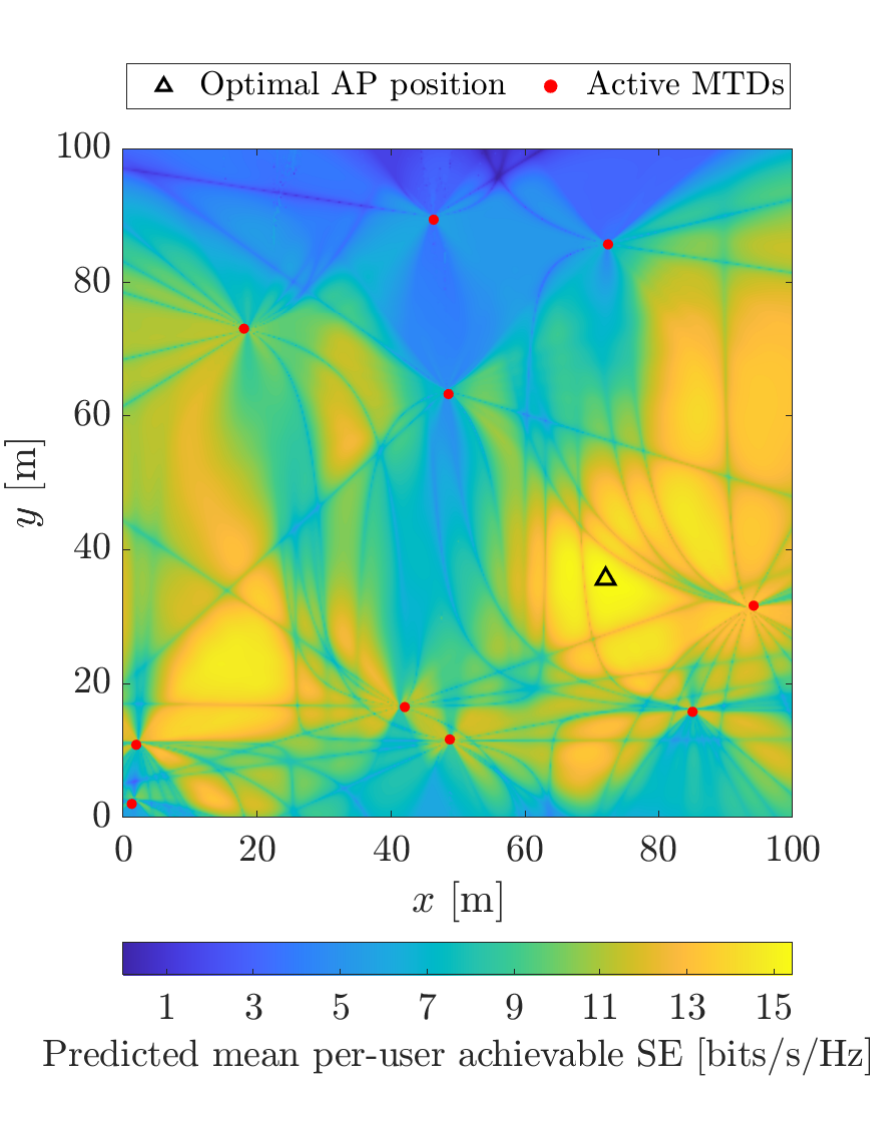}
    \vspace{-0.25cm}
    \caption{Objective function of the MAA, for $L_A=L_B=100$ m, $M=16$, and $K=10$.}
    \label{objectiveFunction_Movable}
\end{figure}

\par Considering a single network realization, i.e., a single set of locations of $K$ active MTDS, we numerically evaluate the objective functions to be optimized. In Fig. \ref{objectiveFunction_Rotary}, we show the objective function for the case of a RAA, which is the predicted mean per-user achievable SE versus the rotation of the AP. Moreover, Fig. \ref{objectiveFunction_Movable} shows the objective function for the case of a MAA, which is the predicted mean per-user achievable SE for all the points of the square coverage area. Note that both objective functions present several local minimum and maximum points. Thus, it is not possible to obtain the optimal points utilizing, for instance, convex optimization techniques \cite{boyd2004}. For this reason, in order to obtain the optimal rotation and/or position of the array, we employ PSO\footnote{The obtained solution is not proven to be optimal, but PSO has been used in numerous non-convex optimization problems showing near optimal performance.}, which will be presented in the next subsection.

\begin{table}[t]
    \centering
    \caption{Parameters of the PSO Algorithm}
    \begin{tabular}{l l l l}
        \toprule
        \textbf{Symbol} & \textbf{Parameter} & \textbf{Symbol} & \textbf{Parameter} \\
        \midrule
        $f(\cdot)$ & Objective function & $c_2$ & Social constant \\
        $x_i$ & \makecell[l]{Position of the $i$-th\\particle} & $p_{b,i}$ & \makecell[l]{Personal best of\\the $i$-th particle} \\
        $v_i$ & \makecell[l]{Velocity of the $i$-th\\particle} & $g_b$ & Global best \\
        $w$ & Inertial weight & $r_1, r_2$ & \makecell[l]{Random numbers\\between 0 and 1}\\
        $c_1$ & Cognitive constant & & \\
        \bottomrule
    \end{tabular}    
    \label{parametersPSO}
\end{table}

\subsection{Particle Swarm Optimization}

\par PSO \cite[Ch. 16]{engelbrecht2007} is an optimization algorithm highly effective for tackling problems where discovering the global maximum or minimum of a function is challenging. This algorithm operates with a population of candidate solutions, referred to as agents or particles, which are moved within the search space based on their current positions and velocities. Each particle's trajectory is influenced by its own best-known position, as well as the global best-known position in the search space. These local and global best positions are updated with each iteration, aiming to direct the swarm of particles towards the optimal solution.

\par The PSO was first introduced in \cite{kennedy1995}, designed to simulate social behaviors such as the motion in bird flocks or fish schools. It has been applied to a variety of optimization problems in communication systems, such as optimal deployment, node localization, clustering, and data aggregation in wireless sensor networks \cite{kulkarni2011}. Additionally, it has been utilized in antenna design to achieve a specific side-lobe level or to determine the positions of antenna elements in a non-uniform array. In communication systems, it has also proven useful in computing the optimal precoding vector to maximize the throughput of a MU-MIMO system \cite{shu2009}, optimizing scheduling in the downlink of MU-MIMO systems \cite{hei2009}, and initializing channel estimates for MIMO-OFDM receivers that simultaneously perform channel estimation and decoding~\cite{knievel2011}.

\par The parameters of the PSO algorithm are listed in Table \ref{parametersPSO}. Moreover, a pseudo-code for the PSO algorithm is listed in Algorithm \ref{PSO}. The inertial weight $w$ controls the particle's tendency to continue in its current direction. Parameters $c_1$ and $c_2$ are the acceleration coefficients, and controls the influence of the personal and global best positions, respectively. The termination criterion might be a pre-determined maximum number of iterations, a certain threshold of the objective function $f(\cdot)$, or any other criteria related to the optimization problem.

\begin{algorithm}[t]
\caption{Particle Swarm Optimization for \(x\), \(y\), and \(\theta\)}
\label{PSO}
\begin{algorithmic}[1]
\State Initialize swarm:
\For{each particle $i$ in the swarm}
    \State Initialize position $\textbf{p}_i=(x_i, y_i, \theta_i)$ randomly
    \State Initialize velocity $\textbf{v}_i=(v_{x_i}, v_{y_i}, v_{\theta_i})$ randomly
    \State Initialize personal best position $\textbf{p}_{b,i}=(x_{b,i}, y_{b,i}, \theta_{b,i})$ to the initial position
\EndFor
\State Initialize global best position $\textbf{g}_b=(x_{gb}, y_{gb}, \theta_{gb})$ to the best initial particle position
\While{Stopping criterion is not met}
    \For{each particle $i$ in the swarm}
        \For{each dimension $(x, y, \theta)$}
            \State Update velocity:
            \State $v_{x_i} = w v_{x_i} + c_1 r_1 (x_{b,i} - x_i) + c_2 r_2 (x_{gb} - x_i)$
            \State $v_{y_i} = w v_{y_i} + c_1 r_1 (y_{b,i} - y_i) + c_2 r_2 (y_{gb} - y_i)$
            \State $v_{\theta_i} = w v_{\theta_i} + c_1 r_1 (\theta_{b,i} - \theta_i) + c_2 r_2 (\theta_{gb} - \theta_i)$
        \EndFor
        \State Update position:
        \State $x_i = x_i + v_{x_i}$
        \State $y_i = y_i + v_{y_i}$
        \State $\theta_i = \theta_i + v_{\theta_i}$
        \If {$f$($\textbf{p}_i$) is better than $f$($\textbf{p}_{b,i}$)}
            \State Update personal best: $\textbf{p}_{b,i}=(x_i, y_i, \theta_i)$
            \If {$f$($\textbf{p}_{b,i}$) is better than $f$($\textbf{g}_b)$}
                \State Update global best: $\textbf{g}_b=(x_{b,i}, y_{b,i}, \theta_{b,i})$
            \EndIf
        \EndIf
    \EndFor
\EndWhile\\
\Return global best position $\textbf{g}_b=(x_{gb}, y_{gb}, \theta_{gb})$
\end{algorithmic}
\end{algorithm}

\section{Numerical Results}
\label{numericalResults}

\par In this section, we present Monte Carlo simulation results to compare the performance achieved by the four different types of APs studied in this work.

\subsection{Simulation Parameters}

\par The power attenuation due to the distance (in dB) is modelled using the log-distance path loss model as
\begin{equation}
    \beta_{k}=-L_0-10\eta\log_{10}\left(\dfrac{d_{k}}{d_0}\right),
\end{equation}
where $d_0$ is the reference distance in meters, $L_0$ is the attenuation owing to the distance at the reference distance (in dB), $\eta$ is the path loss exponent and $d_{k}$ is the distance between the $k$-th device and the AP in meters. The attenuation at the reference distance is calculated using the Friis free-space path loss model and given by
\begin{equation}
    L_0=20\log_{10}\left(\dfrac{4\pi d_0}{\lambda}\right),
\end{equation}
where $\lambda=c/f_c$ is the wavelength in meters, $c$ is the speed of light and $f_c$ is the carrier frequency, as already defined in $^2$.

\par Unless stated otherwise, the values of the simulation parameters adopted in this work are listed in Table \ref{tableParameters}. We assume far-field propagation conditions between all the APs and all the MTDs (please refer to Appendix \ref{appendixFarField}). Moreover, the adopted parameters for the PSO algorithm are listed in Table \ref{parametersPSO_simulation}. Considering the selected values of $M$ and $h_{\text{AP}}$, the communication links between the AP and any device experience far-field propagation conditions.

\par The noise power (in Watts) is given by $\sigma^2_n=N_0BN_F$, where $N_0$ is the power spectral density of the thermal noise in W/Hz, $B$ is the signal bandwidth in Hz, and $N_F$ is the noise figure at the receivers. For the computation of the correlation matrices $\textbf{R}_k,\;\forall k$, we consider $N=6$ scattering clusters, $\psi_{k,n}\sim\mathcal{U}[\phi_k-40\degree,\phi_k+40\degree]$, and $\sigma_{\psi}=5\degree$ \cite{ozdogan2019}.

\begin{table}[]
    \centering
    \caption{Simulation parameters \cite{ngo2017,ozdogan2019,channelModelIndoor}.}
    \begin{tabular}{l l l}
        \toprule
        \textbf{Parameter} & \textbf{Symbol} & \textbf{Value}\\  
        \midrule
        Total number of antenna elements & $M$ & 16\\
        Number of active MTDs & $K$ & $10$\\
        Length of the side of the square area & $L_A$ & $100$ m\\
        Uplink transmission power & $p$ & 100 mW\\
        PSD of the noise & $N_0$ & $4\times10^{-21}$ W/Hz\\
        Signal bandwidth & $B$ & 20 MHz\\
        Noise figure & $N_F$ & 9 dB\\
        Length of the pilot sequences & $\tau_p$ & 10 samples\\
        Length of the time slot & $\tau_p$ & 200 samples\\
        Height of the APs & $h_{\text{AP}}$ & 12 m\\
        Height of the UEs & $h_{\text{UE}}$ & 1.5 m\\
        Carrier frequency & $f_c$ & $3.5$ GHz\\
        Normalized inter-antenna spacing & $\Delta$ & $0.5$\\
        Path loss exponent & $\eta$ & $2$\\
        Reference distance & $d_0$ & $1$ m\\
        \bottomrule
    \end{tabular}    
    \label{tableParameters}
    \vspace{0.3cm}
    \centering
    \caption{Values of the parameters of the PSO Algorithm used for the simulations \cite{MathWorks2024}.}
    \begin{tabular}{l l l}
        \toprule
        \textbf{Symbol} & \textbf{Parameter} & Value \\
        \midrule
        $w$ & Inertial weight & $[0.1,1.1]$\\
        $c_1$ & Cognitive constant & $1.49$\\
        $c_2$ & Social constant & $1.49$ \\
        $N_{\text{vars}}$ & Number of variables & $[1,2,3]$ \\
        $|\mathcal{A}|$ & Swarm size & $\min\{100,10N_{\text{vars}}\}$ \\
        $N_{\text{max}}$ & Maximum number of iterations & $200N_{\text{vars}}$ \\
        $N_{\text{stall}}$ & Maximum number of stall iterations & 20\\
        $\epsilon$ & Function tolerance & $10^{-6}$\\
        \bottomrule
    \end{tabular}    
    \label{parametersPSO_simulation}
\end{table}

\subsection{Simulation Results and Discussions}

\begin{figure*}[htbp]
    \centering
    \begin{minipage}{0.45\textwidth}
        \centering
        \includegraphics[width=0.95\textwidth]{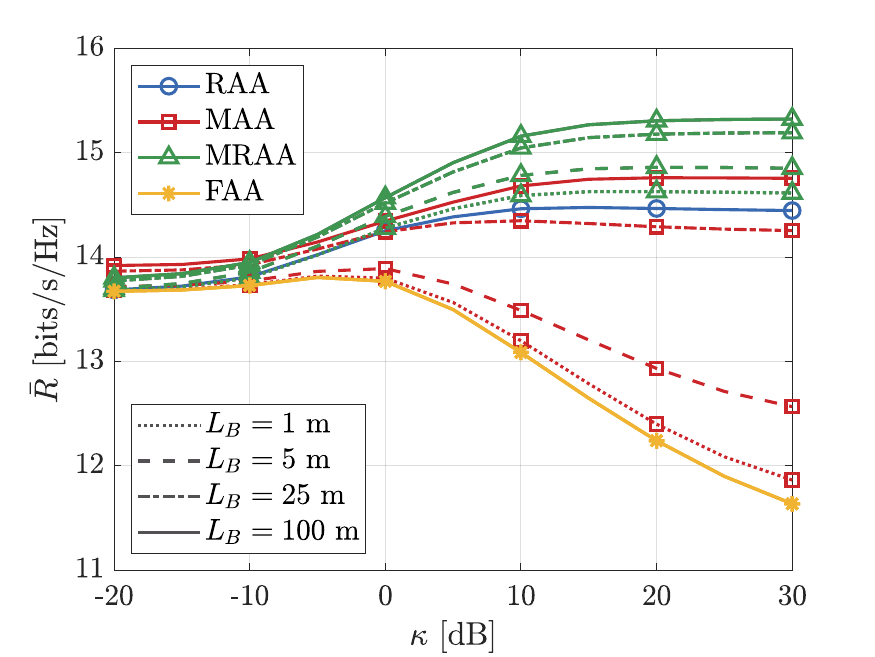}
        \caption{Mean per-user achievable SE versus the Rician factor $\kappa$, for $\sigma_e^2=-10$~dB.}
        \label{plotRicianFactor}
    \end{minipage}
    \hfill
    \begin{minipage}{0.45\textwidth}
        \centering
        \includegraphics[width=0.95\textwidth]{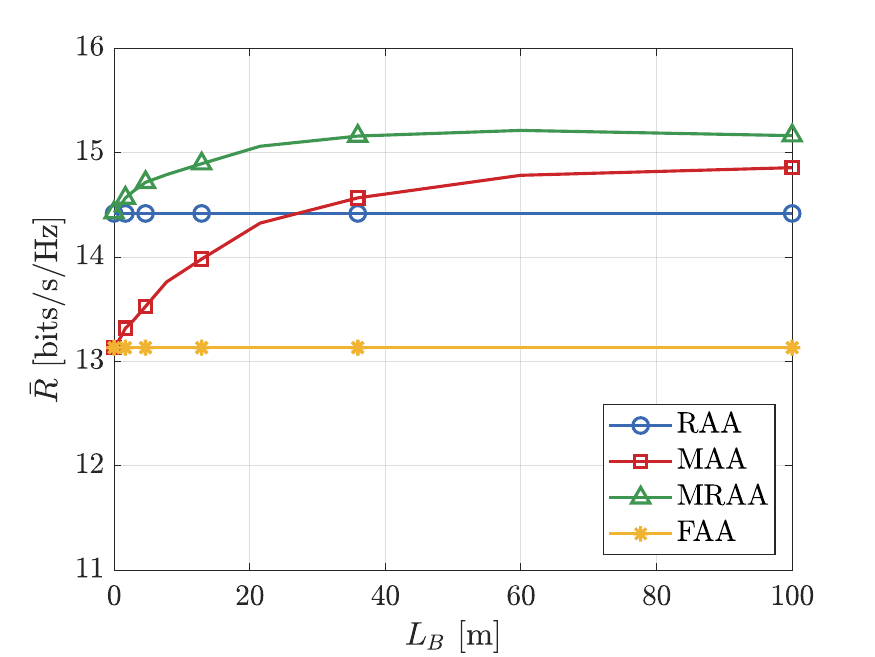}
        \caption{Mean per-user achievable SE versus dimensions of the movement area, for $\kappa=10$ dB and $\sigma_e^2=-10$~dB.}
        \label{plotMovementArea}
    \end{minipage}

    \vskip\baselineskip
    
    \begin{minipage}{0.45\textwidth}
        \centering
        \includegraphics[width=0.95\textwidth]{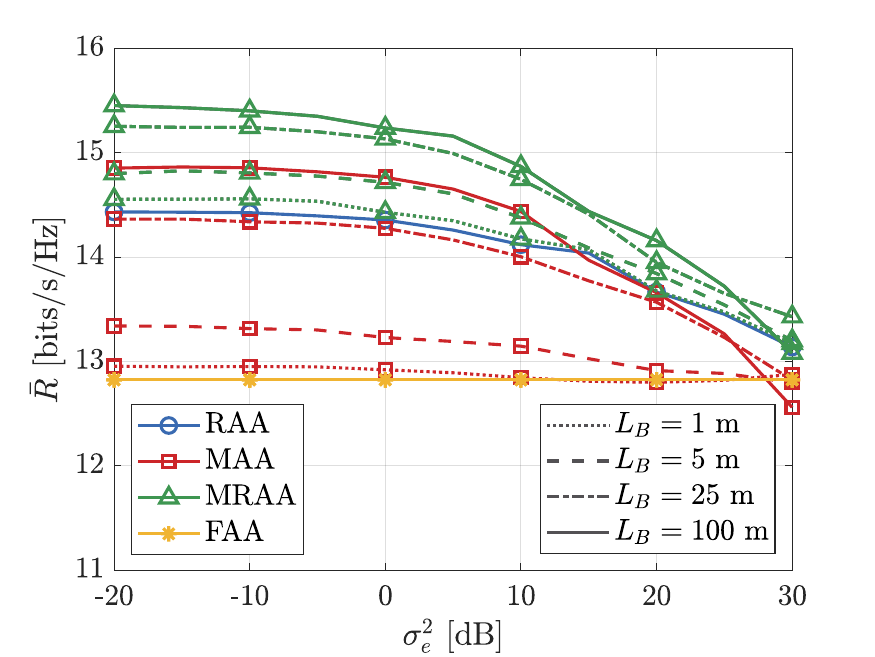}
        \caption{Mean per-user achievable SE versus variance of the localization error considering $\kappa=10$~dB.}
        \label{plotPosError}
    \end{minipage}    
    \hfill    
    \begin{minipage}{0.45\textwidth}
        \centering
        \includegraphics[width=0.95\textwidth]{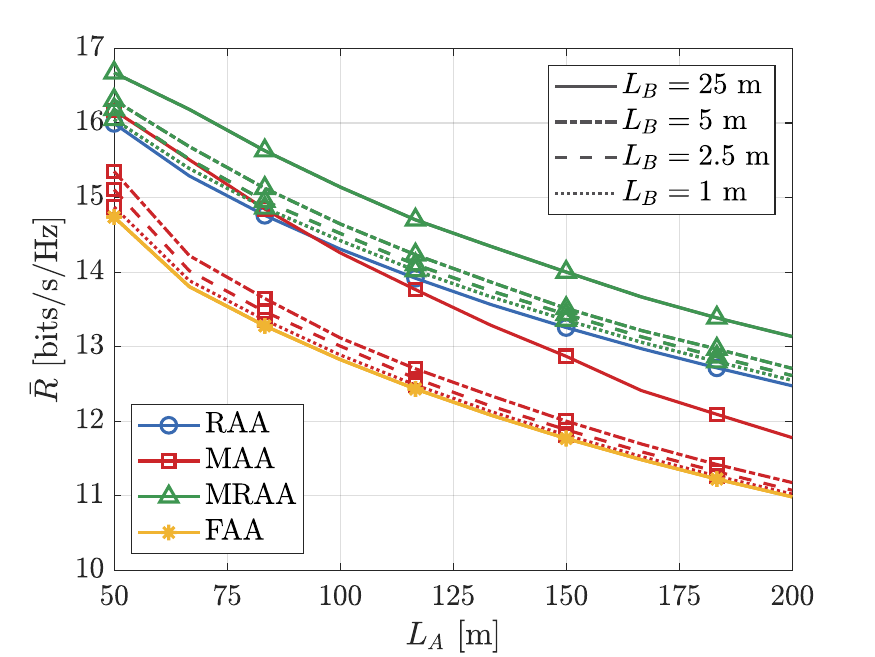}
        \caption{Mean per-user achievable SE versus dimensions of the coverage area, for $\kappa=10$~dB~and~$\sigma_e^2=-10$~dB.}
        \label{plotSizeArea}
    \end{minipage}
    
\end{figure*}

\par We generate average performance results for networks of $K$ devices by averaging the per-user mean achievable SE over multiple network realizations. In other words, the numerical results correspond to the expected SE performance gains for networks of $K$ devices in the considered setup. For each network realization, the achievable SE of the $K$ devices is obtained by averaging over several channel realizations, i.e. distinct realizations of the channel matrix $\textbf{H}$. For each network realization, the locations of the MTDs are fixed and uniformly randomly distributed in the coverage area, i.e., $x_k,y_k\sim\mathcal{U}[0,L_A]$. In the case of MAAs, we evaluate the performance achieved with different sizes of the movement area, i.e., different values of $L_B$.  Note that, in practice, the movement area of the MAAs cannot be very large owing to size and costs constraints. Moreover, moving an array over long distances might take a considerable amount of time (which can be longer than the coherence time of the wireless channel\footnote{In this work, we consider a static scenario where the MTDs do not move. However, in dynamic environments where MTDs or people/objects are moving faster than the AP movement, the coherence time of the channel would be shorter than the moving speed of the AP.}), even when high-speed servo motors are utilized. Thus, the numerical results for the case of $L_B=L_A=100$~m, that is, when the movement area covers the whole coverage area, are hypothetical results that represent an ideal scenario and are utilized here solely for benchmark purposes.

\par Fig. \ref{plotRicianFactor} shows the mean per-user achievable SE versus the Rician factor. In the case of an AP equipped with an FAA or a MAA with a very constrained movement area, we observe that the achievable SE decreases with the Rician factor, while it increases with $\kappa$ when we adopt a RAA or a MAA with a large movement area. As shown in \cite{tominaga2024_2}, the correlation among the channel vectors increases with $\kappa$, which affects the performance obtained with ZF in the case of an FAA or MAA with a small movement area. Nevertheless, by rotating and/or moving the antenna array over a larger area, the AP can find an optimal position and/or rotation that reduces the correlation among the channel vectors.
Note that the performance gains obtained with the optimal rotations or movements become very significant when the LoS component is very strong. As expected, the best performance is achieved with the MRAA, since it has three degrees of freedom for the movements. However, it features the most complex and expensive setup. The MAA, which has two degrees of freedom for the movement, outperforms the RAA only when the movement area covers the whole coverage area. It is very interesting to observe that the RAA outperforms the MAA in the case of $L_B=25$~m, which corresponds to a very large movement area.

\par Fig. \ref{plotMovementArea} shows the mean per-user achievable SE versus the dimensions of the movement area for $\kappa=10$ dB, i.e., a situation where the LoS component of the channel vectors is very strong. When $L_B\rightarrow0$ m, the performance obtained with the MAA converges to the performance obtained with the FAA, while the performance obtained with the MRAA converges to the performance obtained with the RAA, as expected. As we increase the size of the movement area, the performance gains obtained with the movement of the antenna array becomes noticeable. We observe again that the MRAA always presents the best performance. When adopting the MAA, we need a considerably large movement area ($L_B>25$~m) in order to achieve the same performance that can be obtained by simply rotating the antenna array. Note also that the performance improvement obtained by increasing the size of the movement area is negligible for $L_B\geq50$~m.

\par Fig. \ref{plotPosError} shows the mean per-user achievable SE versus the variance of the localization error. We observe that all the optimal rotations and movements of the antenna arrays bring noticeable performance improvements even when the accuracy of the localization information is poor. However, the performance gains on the achievable SE when compared to the case of an FAA decay rapidly for $\sigma_e^2\geq10$ dB. We again observe that the best performance is achieved by the MRAA, followed by the MAA and then the RAA. Nevertheless, note that increasing the accuracy of the localization information to sub-cm or mm levels do not yield performance improvements.

\par Finally, Fig. \ref{plotSizeArea} shows the mean per-user achievable SE versus the dimensions of the coverage area, for $M=16$ and $K=10$. As expected, the achievable SE decreases with $L_A$ for any of the considered setups due to the increased path-losses. We also observe that rotating and/or moving the antenna array always improves the mean per-user achievable SE compared to the case of an FAA. Nonetheless, when the movement area is very constrained, specifically $L_B=1$~m, the performance improvement obtained by moving the array on the horizontal plane is very small. Therefore, a larger movement area is essential to achieve substantial improvements in spectral efficiency. It is noteworthy that the RAA always outperforms the MAAs with $L_B=\{1,2.5,5\}$~m, and it can even outperform the MAA with $L_B=25$~m when $L_A>95$~m, which corresponds to a setup with a large movement area. This finding demonstrates that rotating the array provides greater improvements in the mean per-user achievable SE compared to moving the array along the horizontal plane, even when the movement area on the horizontal plane is relatively large.  Finally, note that the best performance is always achieved by the MRAA, which has more degrees of freedom at the cost of the most complex setup.

\par Overall, the optimal movements and rotations of the antenna arrays bring substantial improvements in the mean per-user achievable SE when the LoS components of the channel vectors is strong. When the movement area is not constrained, the best performance is achieved by the MRAA (which has three degrees of freedom for movement), followed by the MAA (two degrees of freedom) and then by the RAA (one degree of freedom). Nevertheless, the movements of the arrays on the horizontal plane directions need to be constrained in practice due to size, costs, complexity and latency limitations. Moving the array along large distances might induce very high latency. When the movements of the array are constrained, the RAAs outperforms them, while simultaneously presenting a significantly lower size, complexity, and deployment and maintenance costs.

\section{Conclusions}
\label{conclusions}

\par In this work, we compared the performance of MAAs, which are able to move on the horizontal plane using two servo motors, cables and slide tracks, with RAAs, which are equipped a single servo motor and that can rotate on its own axis. We also proposed the combination of both schemes into MRAAs, which are arrays that can move on the horizontal plane and also rotate. The optimal position and/or rotation of the arrays is computed based on estimates of the locations of the active MTDs and using PSO. Our numerical results show that the MAAs outperform the RAAs when their movement area is large enough, but at the cost of a bulkier setup with higher maintenance and deployment costs. When the movement area of the arrays is constrained, the RAAs perform better and also correspond to a simpler and cheaper system. All the proposed techniques offer significant performance gains in terms of mean per-user achievable SE when compared to FAAs when the LoS component of the channel vectors is strong, and all the schemes are robust against imperfect location estimates. 

\appendices

\section{Far-Field Propagation Conditions}
\label{appendixFarField}

\par The Fraunhofer distance determines the threshold between the near-field and far-field propagation, and is given by $d_F=2D^2/\lambda$ \cite{sherman1962}, where $D$ is the largest dimension of the antenna array, $\lambda=c/f_c$ is the wavelength, $c$ is the speed of the light and $f_c$ is the carrier frequency. In the case of an ULA with $M$ antenna elements spaced by half-wavelength, the length of the ULA is $D_{\text{ULA}}=(M-1)\lambda/2$.

\par Considering that a device can be located right bellow an AP in an indoor setting, the minimum height of the AP required to ensure far-field propagation conditions for all the devices is given by
\begin{equation}
\begin{split}
    h_{\text{AP}}^{\text{min}} &= d_F + h_{\text{device}}\\
    &= \dfrac{2}{\lambda}\left((M-1)\dfrac{\lambda}{2}\right)^2+h_{\text{device}}\\
    &= \dfrac{\lambda}{2}(M-1)^2+h_{\text{device}}.
\end{split}
\end{equation}

\par Considering $f_c=3.5$~GHz, $M=16$, $h_{\text{device}}=1.5$~m, we obtain $D_{\text{ULA}}=0.64$~m and $d_F=9.64$~m. Thus, the minimum height of the APs is $h_{\text{AP}}^{\text{min}}=11.14$~m.

\bibliographystyle{./bibliography/IEEEtran}
\bibliography{./bibliography/main}

\end{document}